\documentclass[twocolumn,11pt]{article}

\usepackage{graphicx,color}
\usepackage{amssymb,verbatim,hyperref}
\usepackage[numbers]{natbib}
\usepackage{multibib}
\newcites{body}{References}
\usepackage{natbib}
\setcitestyle{super}
\usepackage{mathptmx}
\begin{document}

\baselineskip=5mm

\twocolumn[{
\noindent {\Huge {\color{blue}\textrm{LETTER}}}\\
\noindent\rule{19cm}{0.4pt}\\

{\huge{\bf
{Warm-hot baryons comprise 5-10 per cent of filaments in the cosmic web \\
}}} 

{\large Dominique Eckert$^{1,2}$, Mathilde Jauzac$^{3,4}$, HuanYuan Shan$^5$, Jean-Paul Kneib$^{5,6}$, Thomas Erben$^7$, Holger Israel$^3$, Eric Jullo$^6$, Matthias Klein$^7$, Richard Massey$^3$, Johan Richard$^8$, C\'eline Tchernin$^1$}\\

{\small
\noindent $^1$Department of Astronomy, University of Geneva, Ch. d'Ecogia 16, 1290 Versoix, Switzerland\\
$^2$INAF - IASF Milano, Via E. Bassini 15, 20133 Milan, Italy\\
$^3$Institute for Computational Cosmology, Department of Physics, Durham University, South Road, Durham DH1 3LE, UK\\
$^4$Astrophysics and Cosmology Research Unit, School of Mathematical Sciences, University of KwaZulu-Natal, Durban 4041, South Africa\\
$^5$Laboratoire dÕAstrophysique, Ecole Polytechnique F\' ed\' erale de Lausanne (EPFL), Observatoire de Sauverny, CH-1290 Versoix, Switzerland\\
$^6$Aix Marseille Universit\'e, CNRS, LAM (Laboratoire d'Astrophysique de Marseille) UMR 7326, 13388, Marseille, France\\
$^7$Argelander-Institut fŸr Astronomie, Auf dem H\"ugel 71, D-53121, Bonn, Germany\\
$^8$CRAL, Observatoire de Lyon, Universit\'e Lyon 1, 9 Avenue Ch. Andr\'e, F-69561 Saint Genis Laval Cedex, France}

\begin{abstract}
\textbf{Observations of the cosmic microwave background indicate that baryons account for 5\% of the Universe's total energy content\citebody{planck15_13}. In the local Universe, the census of all observed baryons falls short of this estimate by a factor of two\citebody{fukugita98,cen99}. Cosmological simulations indicate that the missing baryons might not have condensed into virialized haloes, but reside throughout the filaments of the cosmic web (where matter density is larger than average) as a low-density plasma at temperatures of $\mathbf{10^5-10^7}$ kelvin, known as the warm-hot intergalactic medium\citebody{cen99,dave01,shull12,branchini09}. There have been previous claims of the detection of warm baryons along the line of sight to distant blazars\citebody{fang07,buote09,zappacosta10,nicastro13} and of hot gas between interacting clusters \citebody{kull99,scharf00,zappacosta02,werner08}. These observations were, however, unable to trace the large-scale filamentary structure, or to estimate the total amount of warm baryons in a representative volume of the Universe. Here we report X-ray observations of filamentary structures of gas at $\mathbf{10^7}$ kelvin associated with the galaxy cluster Abell 2744. Previous observations of this cluster\citebody{ibaraki14} were unable to resolve and remove coincidental X-ray point sources. After subtracting these, we reveal hot gas structures that are coherent over scales of 8 mergaparsecs. The filaments coincide with over-densities of galaxies and dark matter, with 5-10\% of their mass in baryonic gas. This gas has been heated up by the cluster's gravitational pull and is now feeding its core. Our findings strengthen evidence for a picture of the Universe in which a large fraction of the missing baryons reside in the filaments of the cosmic web.}\\
\end{abstract}
}]

\smallskip

Abell 2744 is a massive galaxy cluster (containing a total mass of $\sim1.8\times10^{15} M_\odot$ inside a radius of 1.3 Mpc\citebody{merten11}) at a redshift of 0.306\citebody{boschin06,owers11}. In its central regions, the cluster exhibits a complex distribution of dark and luminous matter, as inferred from X-ray and gravitational lensing analyses\citebody{merten11,owers11,kempner04}. Spectroscopic observations indicate large variations in the line-of-sight velocity of different regions\citebody{boschin06,owers11}. Together, these observations reveal that the cluster is currently experiencing a merger of at least four individual components, supporting the hypothesis that Abell 2744 may be an active node of the cosmic web. 

In December 2014, we obtained a 110 ks observation of the cluster by the XMM-Newton X-ray observatory, covering the core and its surroundings out to a radius of $\sim 4 h_{70}^{-1}$ Mpc ($h_{70}=H_0/(\mbox{70 km s}^{-1}\mbox{ Mpc}^{-1})$). We extracted a surface-brightness image of the observation, subtracting a model for the instrumental background and accounting for variation of the telescope efficiency across the field of view. Figure 1 shows the resulting surface-brightness image in the 0.5-1.2 keV band obtained combining the data from the three detectors of the European Photon Imaging Camera (EPIC) on board XMM-Newton. X-ray point sources were masked and the data were adaptively smoothed to highlight the diffuse emission. 

\begin{figure}
\resizebox{\hsize}{!}{\includegraphics{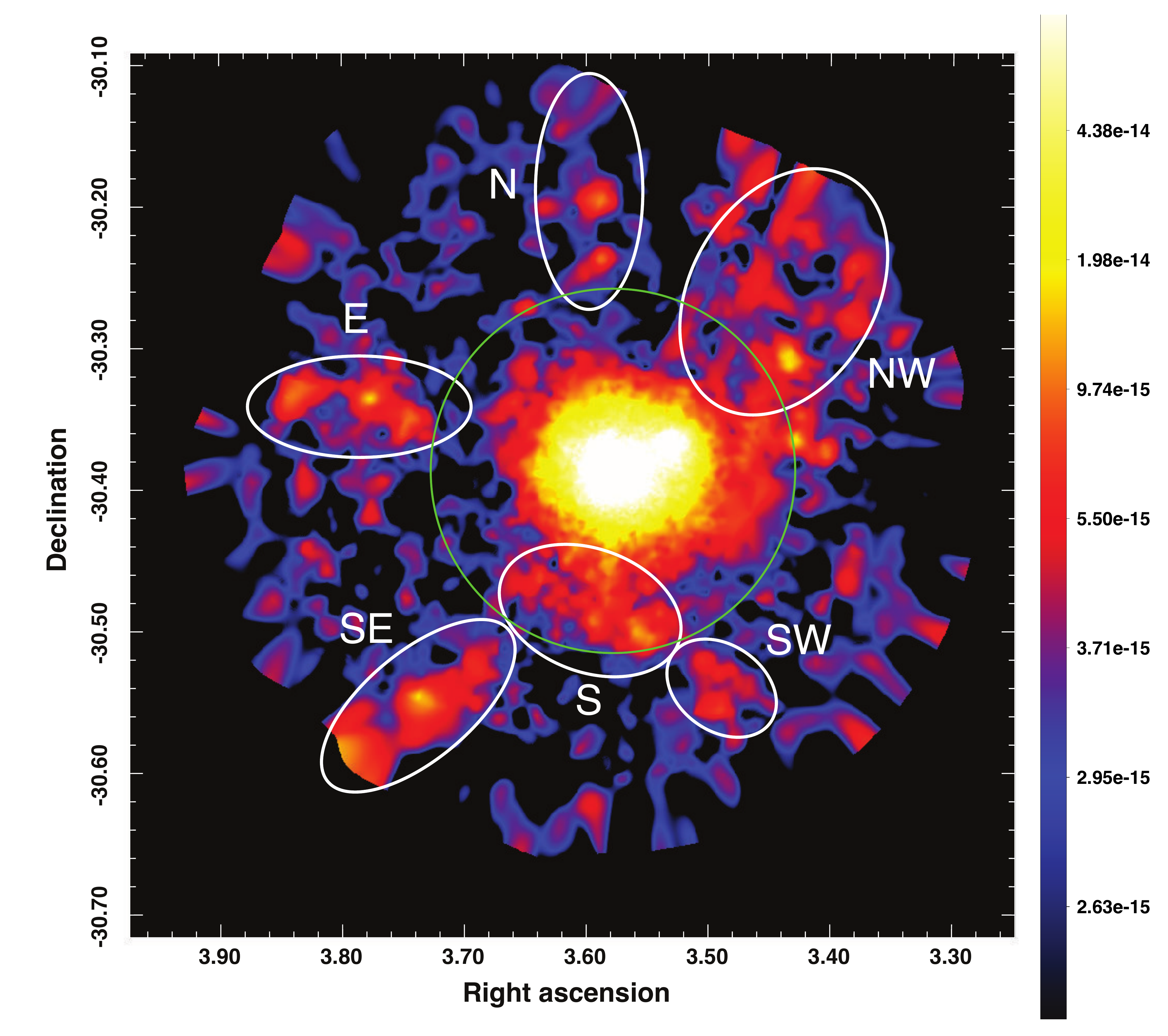}}
\textbf{Figure 1: Map of the hot gas in and around the galaxy cluster Abell 2744.} Shown is the XMM-Newton/EPIC surface-brightness image of the galaxy cluster Abell 2744 in the 0.5-1.2 keV band. The colour bar indicates the brightness in units of ergs cm$^{-2}$ s$^{-1}$ arcmin$^{-2}$. The green circle shows the approximate location of the virial radius $R_{\rm vir} \sim 2.1 h_{70}^{-1}$ Mpc. The white ellipses highlight the position of diffuse structures discovered here.
\end{figure}

The high sensitivity achieved during this observation, thanks to a minimal number of solar flares, allowed us to identify several previously unreported features. Near the virial radius of the cluster ($\sim2 h_{70}^{-1}$ Mpc) and beyond, several high significance ($>6\sigma$) regions of diffuse emission are detected and appear to be connected to the cluster core. To confirm this connection, we extracted the X-ray emissivity profile of the cluster by masking the regions of excess emission, and compared the resulting profile with the emissivity profile in the sectors encompassing the filamentary structures (see Extended Data Figure 1). Although the emissivity of the cluster falls below the detectable level at $\sim2 h_{70}^{-1}$ Mpc from the cluster center, we observe significant emission in sectors extending continuously to the edge of the XMM-Newton field of view, that is, roughly at $4 h_{70}^{-1}$ Mpc in projection from the core. This shows that the detected features are very extended and not caused either by the superposition of unresolved point sources or by individual group-scale haloes. These structures are not visible at higher energies (2-7 keV), in contrast with the cluster core. This suggests that the gas observed in the structures is cooler than that of the central regions.\\

\begin{figure}
\resizebox{\hsize}{!}{\includegraphics{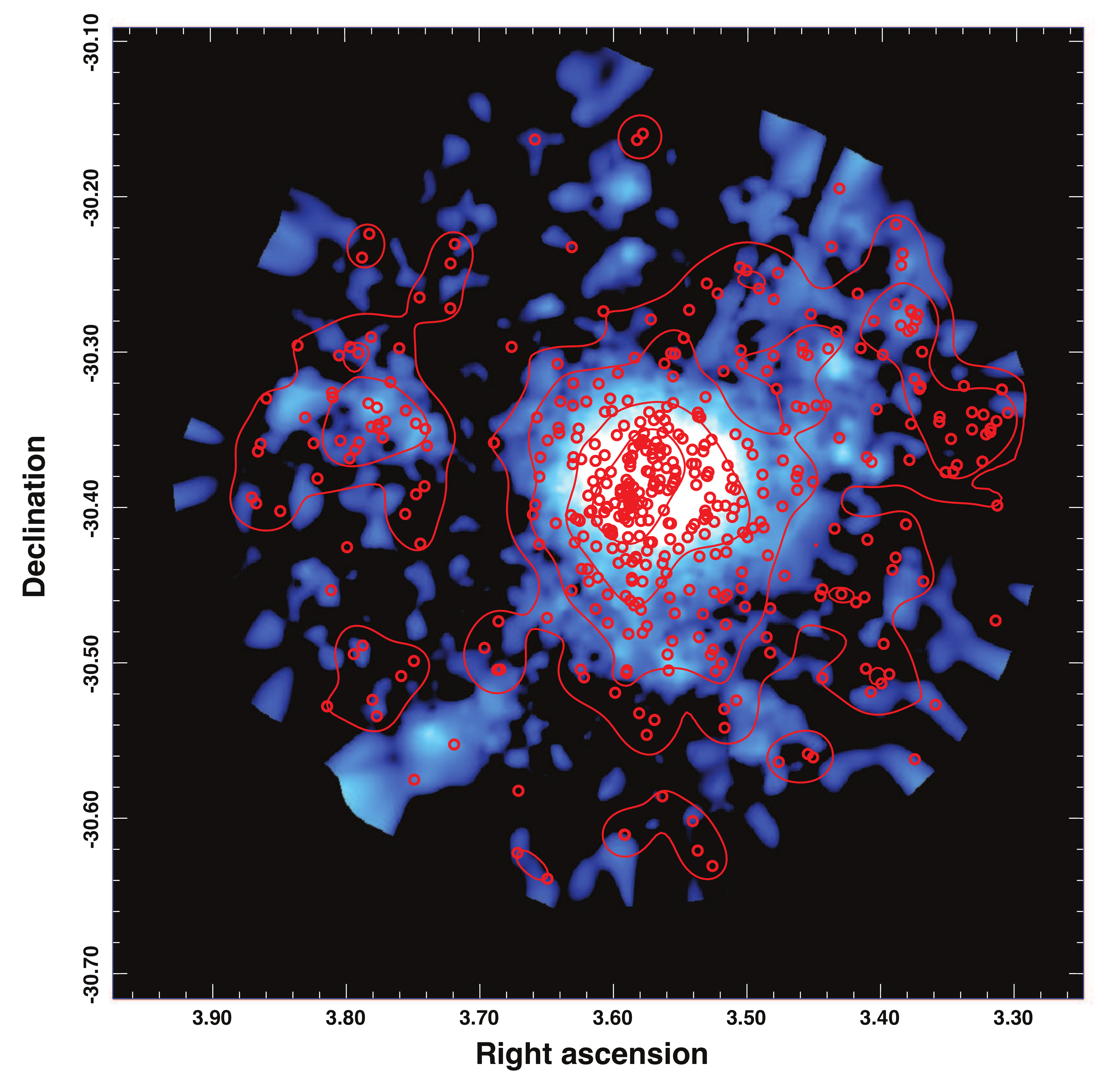}}
\textbf{Figure 2: Comparison between the distribution of hot gas and galaxies in the region surrounding Abell 2744.} Shown is the XMM-Newton image of Abell 2744 (same data as Fig. 1); also shown are the positions of member galaxies with spectroscopic redshift within $\pm5,000$ km s$^{-1}$ of the cluster mean \citebody{owers11} (red dots); red curves show galaxy number density contours.
\end{figure}

To identify the structures detected in X-rays, we used a collection of published spectroscopic redshifts within the XMM-Newton field of view. Spectroscopic redshifts are available for 1,500 galaxies in the field\citebody{boschin06,owers11}. We selected galaxies with velocities falling within $\pm 5,000$ km/s of the cluster mean to capture the cluster and its accretion region in their entirety. In Fig. 2 we show the XMM-Newton brightness image together with the position of selected cluster members and galaxy density contours. Concentrations of cluster galaxies are found coincident with the four hot-gas filamentary structures labeled E, S, SW and NW in Fig. 1. Conversely, structure N corresponds to a background galaxy concentration at $z\approx0.45$, whereas the galaxies associated with the SE substructure exhibit a substantial velocity difference of -8,000 km/s compared to the cluster core. This velocity difference corresponds to a large projected distance from the cluster, which indicates that, although it is part of the same superstructure, this system is probably not interacting with the main cluster. We therefore consider the association of the SE structure with the Abell 2744 complex as tentative and ignore it for the remainder of the analysis. As a result, we only associate structures E, S, SW and NW with the accretion flow towards Abell 2744. Structures S+SW and NW have already identified as galaxy filaments on the basis of the galaxy distribution\citebody{braglia07,ibaraki14}. The average redshift of the galaxies in the E, S and NW structures is consistent with that of the main cluster (see Table 1), indicating that these filamentary structures are oriented close to the plane of the sky. \\

To map the distribution of total mass around the cluster, we measured the weak and strong gravitational lensing of background galaxies visible in wide-field optical images from ground-based telescopes and ultra-deep Hubble Space Telescope (HST) imaging of the cluster core\citebody{jauzac14b}. Our identification of cluster member galaxies utilizes a photometric galaxy catalogue based on Canada-France-Hawaii Telescope (CFHT) data in the i' optical wavelength band and deep, archival data from the Wide-Field Imager (WFI) on the ESO 2.2-m telescope in the B, V and R bands. We selected cluster members and background galaxies using their colours in the BVRiÕ wavelength bands\citebody{israel10}, and used the shear signal measured from a combination of HST and CFHT images for the weak lensing analysis. We used both a simple inversion method and a combined parametric and non-parametric optimization to reconstruct the weak lensing signal. We found that all the substructures identified by XMM-Newton coincide with peaks in the matter distribution, as shown in Figure 3. We then used the weak lensing information to infer an estimate of the mass of the structures detected in X-rays. The total mass within the identified substructures is given in Table 1. Given that dark matter dominates the total mass budget, we conclude that the structures reported here correspond to overdensities in both the baryon and dark matter distribution. \\

Wide-field galaxy redshift surveys have shown that the large-scale distribution of matter in the Universe is not homogeneous\citebody{bond96,yess96}. Instead, matter tends to fall together under the action of gravity into filamentary structures, forming the cosmic web\citebody{bond96,springel06}. Galaxy clusters, the largest gravitationally-bound structures in the Universe, form at its nodes, where the matter density is the highest. We therefore associate the structures discovered here with intergalactic filaments and conclude that Abell 2744 is an active node of the cosmic web.\\

\begin{figure}
\resizebox{\hsize}{!}{\includegraphics{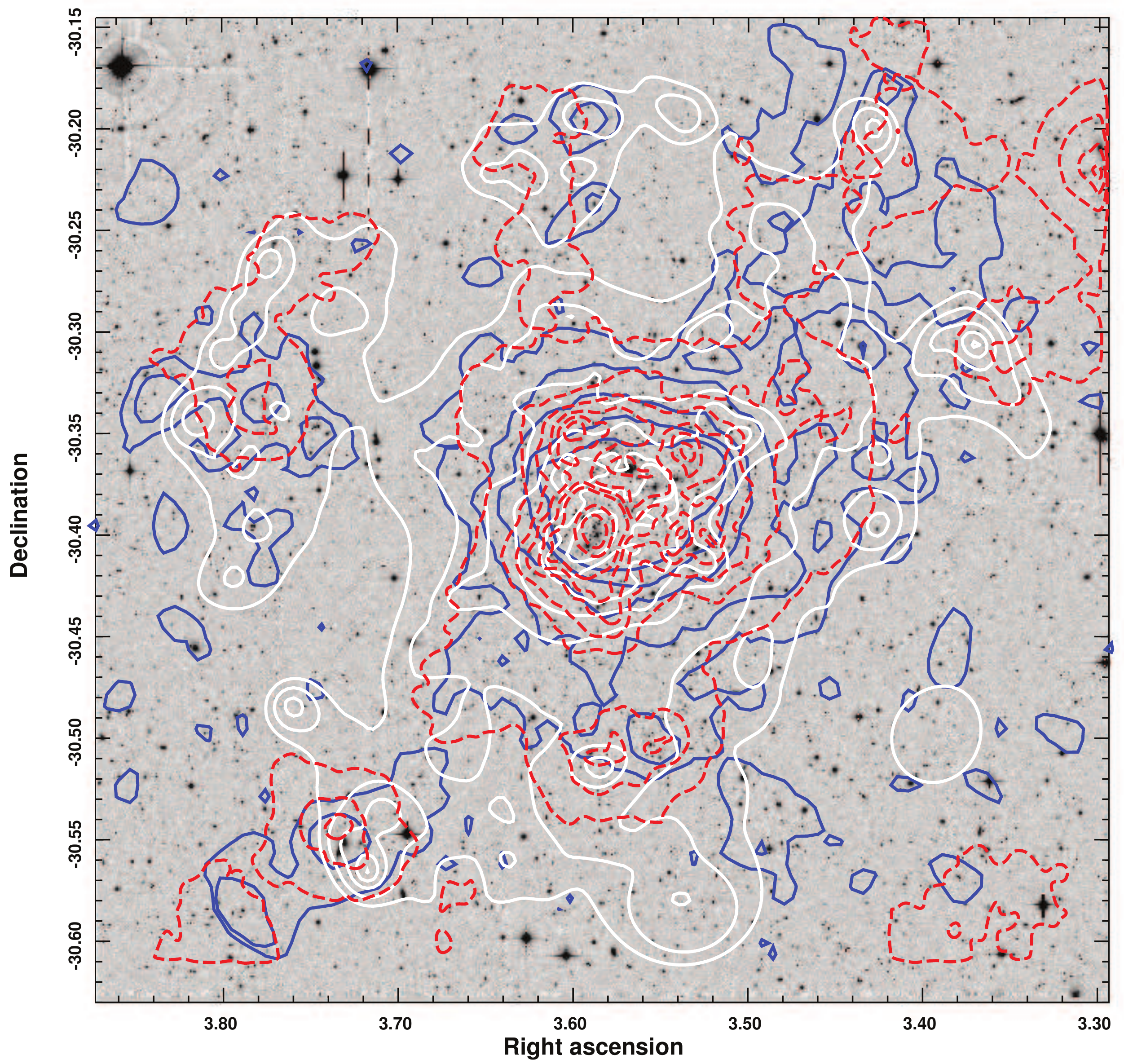}}
\textbf{Figure 3: Hot gas, visible light and total mass in Abell 2744.} Shown is the CFHT image of Abell 2744 and the surrounding large-scale structure. The contours show X-ray isophotes (blue), mass distribution reconstructed from combined strong and weak lensing (white), and optical light (dashed red).
\end{figure}

We estimated the plasma temperature in all the filaments highlighted in Fig. 1 by extracting their X-ray spectra and fitting them with a thin-plasma emission model. The gas in the structures has a typical density of about $10^{-4}$ particles per cm$^3$, corresponding to overdensities of $\sim200$ compared to the mean baryon density\citebody{takei07}. Approximating the geometry of the filaments as segments of cylinders, we estimate the total gas mass enclosed within the filaments to be considerable ($\sim4\times10^{13} M_\odot$). Given the mass within the filaments obtained from weak lensing, we estimate a gas fraction between 5 and 15\% for the various substructures, depending on the adopted mass reconstruction method (see Table 1), which represents a significant fraction of the Universal baryon fraction of 15\%\citebody{planck15_13}. The plasma temperature is in the range $(10-20)\times10^6$ K for the various filaments (see Table 1). This is substantially less than the virial temperature of the cluster core ($\sim10^8$ K), which indicates that the plasma has not yet virialized within the main dark-matter halo. These gas temperatures and densities correspond to those expected for the hottest and densest parts of the warm-hot intergalactic medium (WHIM)\citebody{cen99,dave01,dolag06,gheller15}. Numerical simulations predict that the bulk of the gas permeating intergalactic filaments should have temperatures in the range $10^{5.5}-10^{6.5}$ K, but the gas in the vicinity of the cluster may have undergone substantial heating caused by adiabatic compression and shock heating. Note also that the temperatures measured here may be significantly overestimated, given that X-ray telescopes are sensitive preferentially to the hottest phase of the expected gas distribution. Overall, these properties support the picture in which a large fraction of the Universe's baryons are located in the filaments of the cosmic web.\\

\begin{table*}
\begin{center}
\begin{tabular}{cccccccc}
\hline
Region & $\langle z\rangle$ & $T$ & $M_{\rm gas}$ & S/N & $M_{\rm tot}$ & S/N & $f_{\rm gas}$\\
 & & [$10^6$ K] & [$h_{70}^{-1}M_\odot$] & X-ray & [$h_{70}^{-1}M_\odot$] & lensing & \\
\hline
\hline
E & $0.308$ & $15\pm2$ & $(3.8\pm0.6)\times10^{12}$ & 15.4 & $(7.9\pm2.8)\times10^{13}$ & 3.1 & $0.05\pm0.02$\\
S & $0.303$ &  $16\pm2$ & $(7.1\pm0.8)\times10^{12}$ & 22.6 & $(9.5\pm2.4)\times10^{13}$ & 6.8 & $0.07\pm0.02$\\
SW & $0.305$ & $8^{+4}_{-2}$ & $(2.0\pm0.4)\times10^{12}$ & 9.6 & $(4.8\pm1.7)\times10^{13}$ & 3.1 & $0.04\pm0.02$\\
NW1 & $0.305$ & $25\pm4$ & $(5.7\pm0.3)\times10^{12}$ & 25.3 & $(9.5\pm2.7)\times10^{13}$ & 5.2 & $0.06\pm0.02$\\
NW2 & $0.305$ & $19\pm2$ & $(1.9\pm0.1)\times10^{13}$ & 25.9 & $(1.2\pm0.3)\times10^{14}$ & 3.3 & $0.15\pm0.04$\\
 \hline
\end{tabular}
\end{center}
\textbf{Table 1: Properties of the filaments discovered in this study.} X-ray and lensing properties of the regions defined in Extended Data Figure 2. Note that because of the uncertainty in the geometry of the filaments the provided gas mass, total mass, and gas fraction should be considered as indicative. The masses reported here were obtained by combining strong and weak lensing. A comparison with weak-lensing-only measurements is provided in Extended Data Table 2.
\end{table*}

{\small
\bibliographystylebody{naturemag}
\bibliographybody{nature}
}

\vspace{0.2cm}
\emph{Acknowledgements:} Based on observations obtained with XMM-Newton, an ESA science mission with instruments and contributions directly funded by ESA Member States and NASA. 
DE thanks F. Vazza, S. Paltani and S. Molendi for fruitful discussions. We thank H. Ebeling, M. Limousin, B. Cl\' ement, H. Atek, D. Harvey, E. Egami, M. Rexroth and P. Natarajan for their help with the writing of the XMM-Newton proposal.
MJ, HI and RM acknowledge support from the UK Science and Technology Facilities Council (grant numbers ST/L00075X/1, ST/H005234/1), the Leverhulme trust (grant number PLP-2011-003) and the Royal Society.
JPK acknowledges support from the ERC advanced grant LIDA and from CNRS.
HYS acknowledges support by a Marie Curie International Incoming Fellowship within the 7th European Community Framework Programme, and NSFC of China under grant 11103011.
TE is supported  by  the  Deutsche  Forschungsgemeinschaft  through  the Transregional Collaborative Research Centre TR 33 - `` The Dark Universe ''.
EJ is supported by CNES.
JR acknowledges support from the ERC starting grant CALENDS.
\vspace{0.3cm}

\noindent {\large \textbf{Author contributions}}

\noindent DE: Lead author, X-ray analysis\\
MJ: Weak and strong lensing analysis\\
HYS: CFHT weak lensing analysis\\
JPK: Principal investigator of the XMM-Newton observation, strong and weak lensing analysis and identification of the red cluster sequence in the photometric data\\
TE: WFI and CFHT data reduction\\
HI: WFI and CFHT data reduction\\
EJ: Weak and strong lensing modeling techniques\\
MK: WFI and CFHT data reduction\\
RM: Weak lensing analysis\\
JR: Strong lensing analysis\\
CT: X-ray analysis\\

\noindent {\large \textbf{Author Information}}

\noindent Reprints and permissions information is available at www.nature.com/reprints. There are no competing financial interests. Correspondence and requests for materials should be addressed to \texttt{Dominique.Eckert@unige.ch}. \\

\newpage

\quad

\newpage

\twocolumn[{\centerline{\textbf{\huge Methods}}
\vspace{0.2cm}}]

{\large \textbf{\emph{Imaging X-ray analysis}}}\\

Abell 2744 was observed by XMM-Newton in late 2014 for a total observing time of 110 ks (PI: J.-P. K.; OBSID 074385). At the redshift of A2744 (0.306), the size of the XMM-Newton field of view corresponds to $8 h_{70}^{-1}$ Mpc. We processed the data using the XMM-Newton Scientific Analysis System (XMMSAS) v14.0. We excluded flaring periods from the event files by creating a light curve for each instrument separately and filtering out the time periods for which the observed count rate exceeded the mean by more than $2\sigma$. The observation was very mildly affected by soft-proton flares, allowing us to reach a flare-free observing time of 96 ks, 97 ks, and 87 ks for EPIC detectors MOS1, MOS2 and pn, respectively. 

We extracted raw images in the 0.5-1.2 keV band for all three EPIC detectors using the Extended Source Analysis Software (ESAS) package\cite{snowden08}. This energy band maximizes the source-to-background ratio and avoids the bright Al and Si background emission lines, while maintaining a large effective area since the collecting power of the XMM-Newton telescopes peaks at 1 keV. Exposure maps for each instrument were created, taking into account the variations of the vignetting across the field of view. A model image of the non X-ray background (NXB) was computed using a collection of closed-filter observations, and was adjusted to each individual observation by comparing the count rates in the corner of the field of view. X-ray point sources were detected using the XMMSAS tool \texttt{ewavelet} and masked during the analysis. Additionally, we used the existing Chandra observations of the cluster\citebody{owers11,kempner04} to detect point sources down to fainter X-ray fluxes ($\sim5\times10^{-16}$ ergs cm$^{-2}$ s$^{-1}$) and mask the corresponding areas. Such a flux threshold for point-source removal corresponds to a resolved fraction of 80\% of the cosmic X-ray background\cite{moretti03}, which is associated with a cosmic variance of about 5\%. This ensures that the extended features reported here are indeed caused by diffuse emission.

We computed surface-brightness images by subtracting the NXB from the raw images and dividing them by the exposure maps. To maximize the signal-to-noise ratio (SNR), we then combined the surface-brightness images of the three EPIC detectors by weighting each detector by its relative effective area. The resulting image was then adaptively smoothed using the XMMSAS tool \texttt{asmooth}, requiring an SNR of 5 for all features above the local background. The total XMM-Newton/EPIC image of A2744 is shown in Fig. 1.

To confirm the presence of the filamentary structures shown in Fig. 1, we compared the surface brightness of the regions inside and outside the filaments. We used the PROFFIT code\cite{eckert11} to extract the surface brightness profile from the surface-brightness peak by masking the sectors corresponding with the filaments, and we compared the masked profile with the surface brightness profile in direction of the filaments, that is, in the sectors including the filaments (position angles $10-70^\circ$, $150-180^\circ$, and $260-300^\circ$ for the NW, E and S filaments, respectively, where $0^\circ$ is the W direction; see Extended Data Fig. 2). In Extended Data Fig. 1 we show the corresponding surface-brightness profiles. When masking the filaments, no statistically significant cluster emission is detected beyond 7 arcmin ($\sim2 h_{70}^{-1}$ Mpc); in the direction of the filaments, a flat surface brightness is observed out to the edge of the field of view ($\sim4 h_{70}^{-1}$ Mpc). The small variations in the amplitude of the surface-brightness profiles indicates that the emission is due to filamentary structures rather than to a collection of infalling clumps. The excess emission produced by the filaments was already noted in \emph{Suzaku} observations of the cluster\citebody{ibaraki14}; the poor angular resolution and narrow field of view of \emph{Suzaku} were however insufficient to separate the filaments from the field and resolve point sources. 

\begin{figure}
\resizebox{\hsize}{!}{\includegraphics{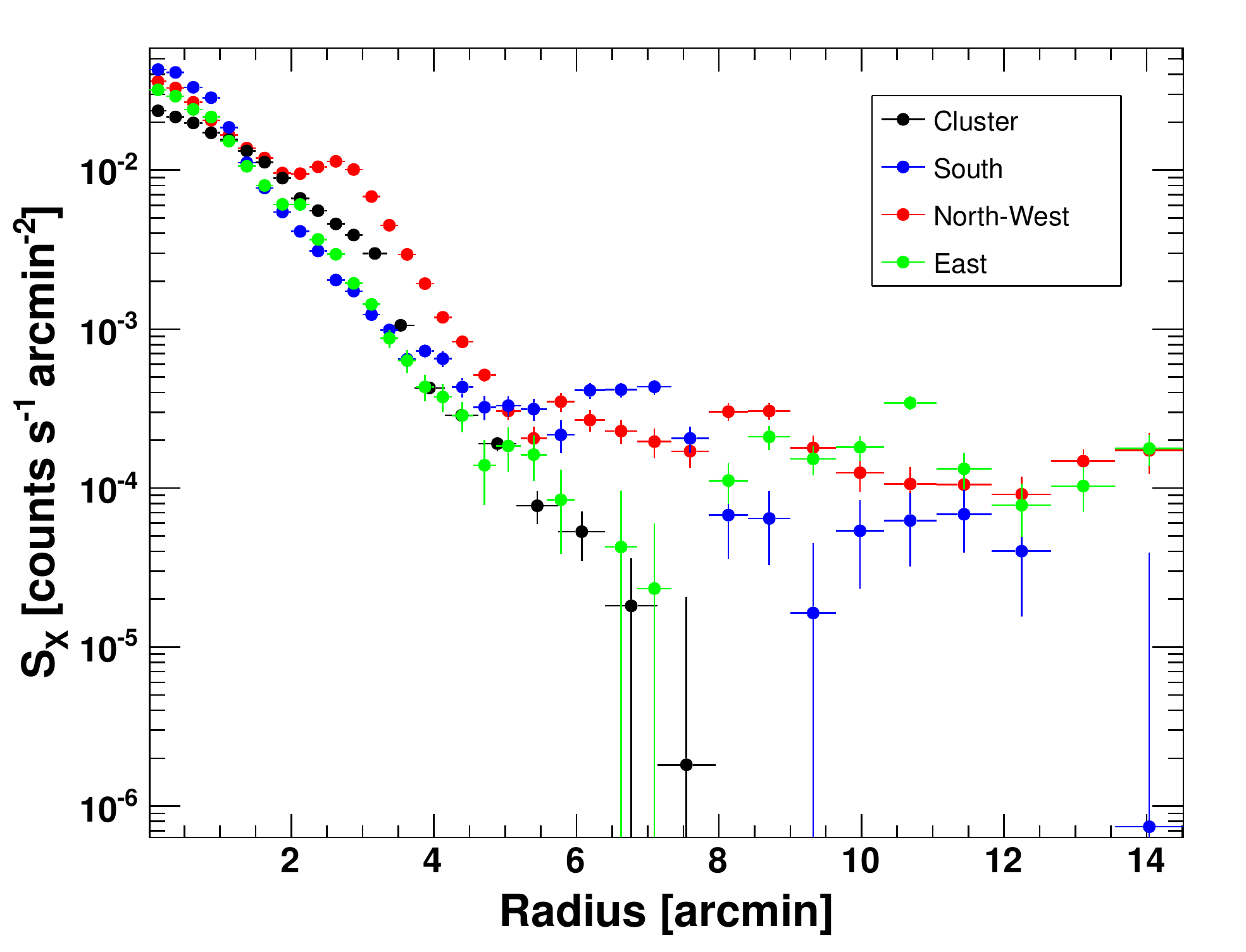}}
\textbf{Extended Data Figure 1: Radial X-ray emissivity profiles in the filaments and in the cluster.} Shown are XMM-Newton/EPIC surface brightness profiles ($S_X$); black, obtained by masking the filaments; colours, surface brightness in the sectors NW (northwest, position angle $10^\circ-70^\circ$), E (east, $150^\circ-180^\circ$), and S (south, $260^\circ-300^\circ$). Uncertainties (error bars) are given at the 1$\sigma$ level.
\end{figure}

For comparison, we extracted radial profiles of galaxy density from spectroscopically-confirmed members\citebody{owers11} in exactly the same sectors. The resulting profiles are shown in Extended Data Fig. 3. We find that beyond the cluster's virial radius the galaxy density is consistently larger in the regions containing the filaments compared to the perpendicular directions, which highlights the association between the structures detected in X-rays and the local galaxy distribution.\\

\begin{figure}
\resizebox{\hsize}{!}{\includegraphics{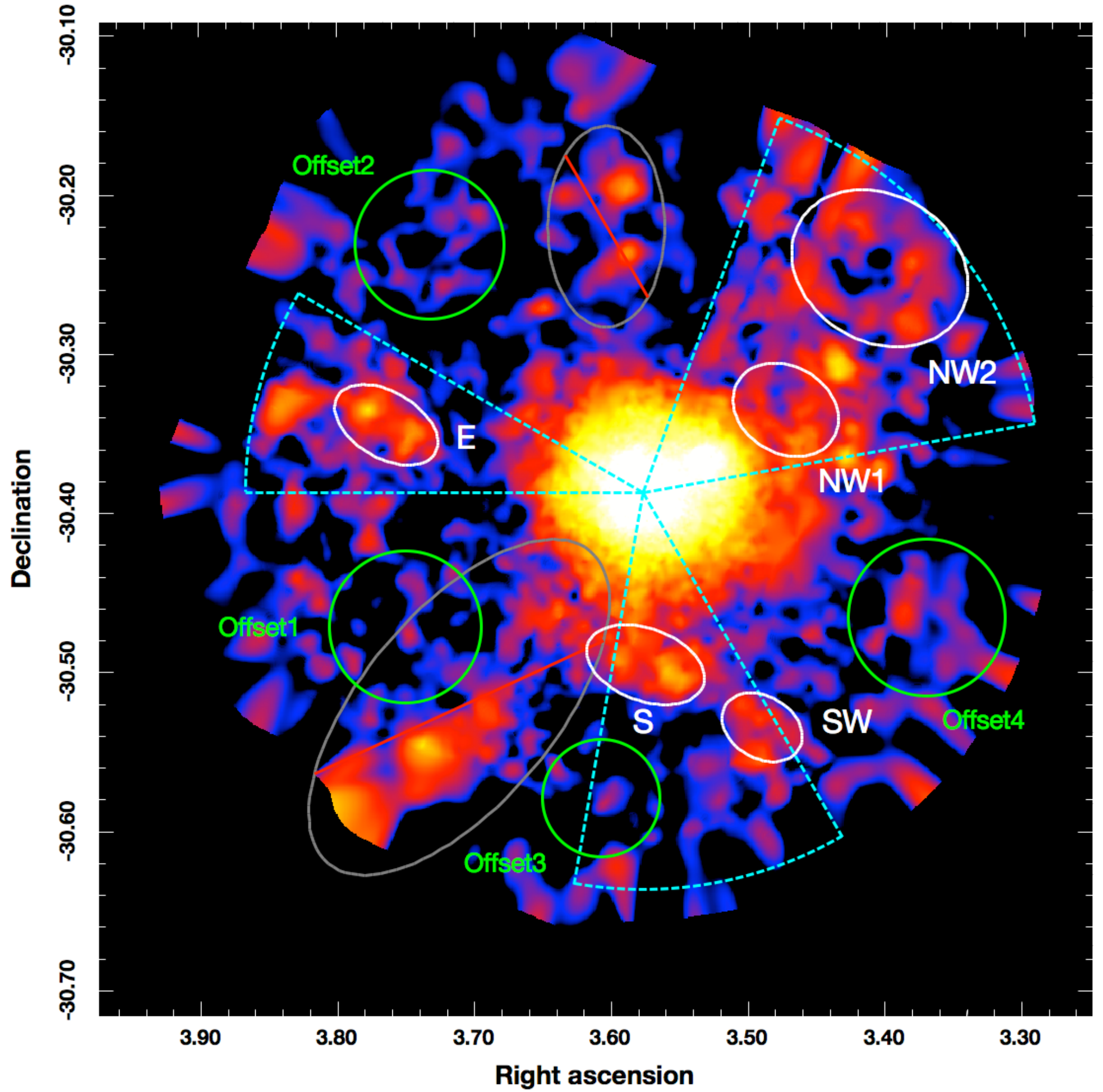}}
\textbf{Extended Data Figure 2: Regions used for the analysis of the thermodynamic properties of the filaments.} The 0.5-2 keV surface brightness is colour coded. Spectra were extracted from the regions indicated as E, S, SW, NW1 and NW2 by the white ellipses. The green circles show the regions labelled as Offset1-4 used to estimate the local background components (see Extended Data Table 1). The dashed cyan sectors show the regions used to extract the radial profiles along the filaments for Extended Data Figs 1, 3 and 5. The grey ellipses show background/foreground structures masked during the analysis (see text).
\end{figure}

{\large \textbf{\emph{Spectral X-ray analysis}}}\\

We performed a spectral analysis of the structures highlighted in Fig. 1. We defined elliptical regions following the X-ray isophotes as closely as possible. In Extended Data Fig. 2 we show the regions used to derive the spectral properties of the filaments. Since the surface brightness of these regions barely exceeds the background level, a detailed modeling of all the various background components is necessary to obtain reliable measurements of the relevant parameters. We adopted the following approach to model the various spectral components\cite{eckert14b}:

\begin{figure}
\resizebox{\hsize}{!}{\includegraphics{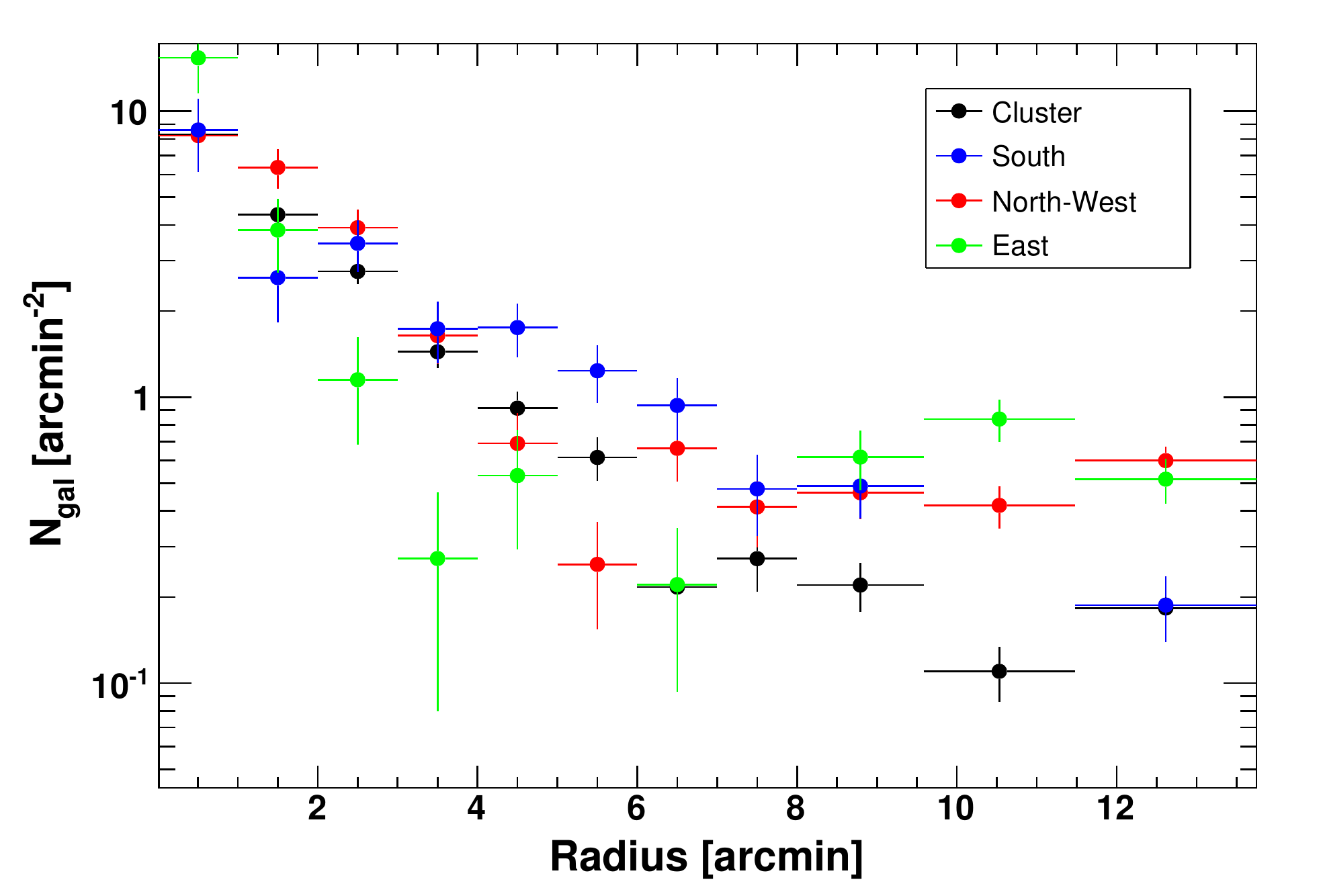}}
\textbf{Extended Data Figure 3: Radial galaxy density profiles in the filaments and in the cluster.} Galaxy density profiles ($N_{\rm gal}$) using spectroscopically confirmed cluster members in sectors encompassing the filaments (same as Extended Data Fig. 1) are compared to the galaxy density of the cluster obtained by masking the filaments (black). Uncertainties (error bars) are given at the 1$\sigma$ level.
\end{figure}

\begin{itemize}
\item \textit{The source:} We modelled the diffuse emission in each region using the thin-plasma emission code APEC\cite{apec}, leaving the temperature and normalization as free parameters. The metal abundance Z was fixed to $0.2Z_\odot$\cite{lm08b}. This component is absorbed by the Galactic column density, which we fixed to the 21cm value ($N_H=1.5\times10^{20}$ cm$^{-2}$ \cite{kalberla}).
\item \textit{The non X-ray background (NXB):} We used closed-filter observations to estimate the spectrum of the NXB component in each region\cite{snowden08}. Instead of subtracting the NXB, we modelled it using a phenomenological model and included it as an additive component in the spectral fitting. This method has the advantage of retaining the statistical properties of the original spectrum. We left the normalization of the NXB component free to vary during the fitting procedure, which allows us to take variations of the NXB level into account. The normalization of the prominent background lines was also left free. Since the observation was very weakly contaminated by soft proton flares, the residual soft proton component can be neglected.
\item \textit{The sky background components:} We used 4 offset regions where no cluster emission is detected (see Extended Data Fig. 2) to measure the sky background components in the field of A2744. We modelled the sky background using a three-component model: \emph{i)} a power law with photon index fixed to 1.46 to model the cosmic X-ray background (CXB); \emph{ii)} a thermal component at a free temperature to estimate the Galactic halo emission; \emph{iii)} an unabsorbed thermal component at 0.11 keV for the local hot bubble. The best-fit spectrum for the Offset1 region is shown in the top-left panel of Extended Data Fig. 4. In Extended Data Table 1 we show the best-fit parameters for our sky background model in the four offset regions. The variation of the parameters from one region to another gives us a handle of the systematic uncertainties associated with the variation of the sky background across the field of view. The main sky component (the CXB) typically varies by $\pm10\%$ across the field. Slightly larger variations ($\sim20\%$) are observed for the foreground components, although it most be noted that the normalizations of the Galactic halo and local bubble components is correlated. The overall values of these parameters agree well with previous measurements of the CXB\cite{deluca04} and the foregrounds\cite{mccammon02}.
\end{itemize}

We note that because of strong Galactic absorption in the far-ultraviolet band and falling effective area in this wavelength range, XMM-Newton is sensitive predominantly to the hottest phase of the gas ($T>10^{6.5}$ K). To test the sensitivity of our observations to cooler plasma, we assumed a differential emission-measure model including gas temperatures in the range $10^{5.5}-10^7$ K and simulated an XMM-Newton spectrum at the same depth as our observation. The resulting spectrum can be well fitted with a single-temperature model at $T=10^{6.8}$ K. This indicates that the temperatures measured here may be substantially overestimated if the plasma is multiphase. 

\begin{table*}
\begin{center}
\begin{tabular}{ccccc}
\hline
Region & CXB & Halo $kT$ & Halo Norm & LB Norm\\
\hline
\hline
Offset 1 &  $(6.26 \pm 0.56)\times10^{-7}$ & $0.297 \pm 0.024$  & $(4.45 \pm 0.60)\times10^{-7}$ & $(1.89 \pm 0.25)\times10^{-6}$ \\
Offset 2 & $(7.03 \pm 0.71)\times10^{-7}$ & $0.368\pm0.095$ & $(2.31\pm0.91)\times10^{-7}$ & $(2.36\pm0.36)\times10^{-6}$\\
Offset 3 & $(6.92\pm0.78)\times10^{-7}$ & $0.311\pm0.034$ & $(5.05\pm0.88)\times10^{-7}$ & $(2.14\pm0.36)\times10^{-6}$\\
Offset 4 & $(7.65\pm0.71)\times10^{-7}$ & $0.283\pm0.036$ & $(3.52\pm0.82)\times10^{-7}$ & $(2.40\pm0.28)\times10^{-6}$\\
\hline
\end{tabular}
\end{center}
\textbf{Extended Data Table 1: Properties of the X-ray background in the A2744 region.} Comparison of X-ray background parameters per square arcminute obtained in regions Offset 1 through Offset 4 (see Extended Data Fig. 2). The units of the column are: photons keV$^{-1}$ cm$^{-2}$ s$^{-1}$ at 1 keV (CXB); keV (Halo $kT$); $\int n_en_H\,dV\times10^{-14}/(4\pi d_A^2(1+z)^2)$ (Halo and Local Bubble normalizations).
\end{table*}

In Extended Data Fig. 4 we show the observed spectra for the five regions defined in Extended Data Fig. 2 together with their best-fit models. Since it is the brightest and most extended, the NW filament was split into two regions (labelled NW1 and NW2) to study the variation of the spectral parameters along a single filament. The resulting parameters are provided in Table 1. To estimate the gas mass within each filamentary structure, we modeled the emission region as a cylinder with length and diameter given by the major and minor axes of the defined ellipses, respectively. We converted the measured normalization into an emission measure, and computed the average gas density assuming constant density in each structure. We estimated the gas mass by integrating the resulting gas density over the volume (see Table 1). We note that given the large uncertainties in the 3D geometry of the filaments, the recovered gas densities and masses should be considered as indicative. Indeed, we tested the effect of adopting different geometries (spheres, ellipsoids) on the recovered gas mass and gas density, and found that the results obtained with the various geometries vary by $\sim30\%$.

\begin{figure*}
\resizebox{\hsize}{!}{\includegraphics{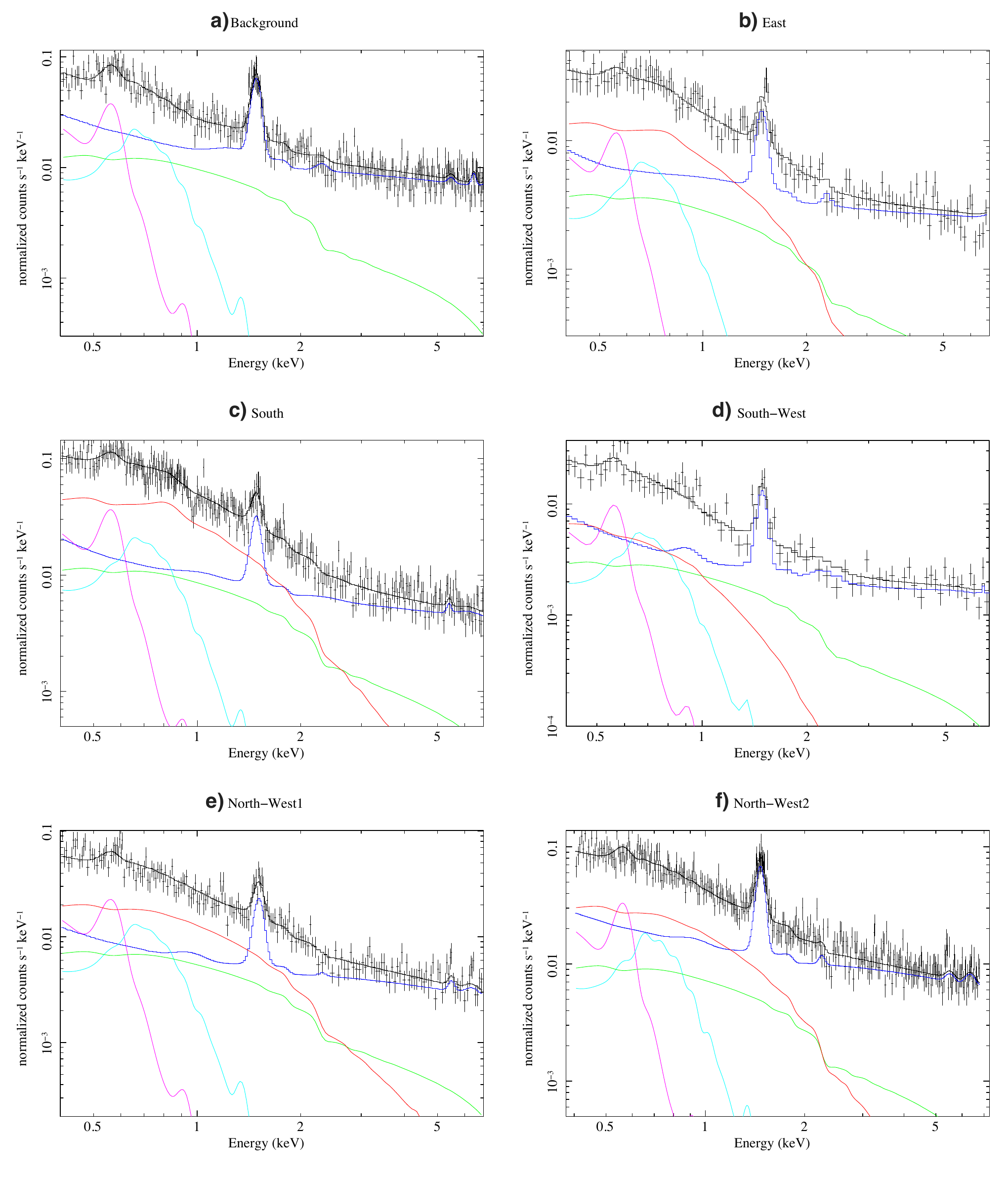}}
\textbf{Extended Data Figure 4: X-ray spectra of the filaments.} XMM-Newton/EPIC-pn spectra for the regions shown in Extended Data Fig. 2. The background region (a) refers to Offset1. The fitting procedure was performed jointly on all EPIC instruments, however only the pn spectra are shown here for clarity. The various coloured lines show fitted contributions from the source (red), the NXB (blue), the CXB (green), the Galactic halo (cyan), and the local hot bubble (magenta).
\end{figure*}

To assess the level of systematic uncertainties in our spectral measurements, we used the spectrum of the SW region, as it is the weakest and thus is the most prone to systematic uncertainties, and let the various sky background and NXB parameters vary within their allowed ranges. We then applied a Markov chain Monte Carlo (MCMC) algorithm to sample the likelihood distribution. The posterior distribution for the measured parameters are then marginalized over the systematic uncertainties associated with the variation of the background components. Through this approach, we found a typical systematic uncertainty of $\sim20\%$ on the gas temperature and $<5\%$ on the emission measure. These values provide an upper limit to the level of systematic uncertainties in the other regions since the intensity of the source relative to the background is higher than for the analysis carried out here.\\

{\large \textbf{\emph{Analysis of ESO and CFHT optical data}}}\\

We used the colours of galaxies in archival optical imaging of the Abell 2744 field to identify members of the cluster and its associated filaments. We constructed a photometric catalogue from observations obtained in the B,V and R filters using the WFI instrument at the ESO 2.2-m telescope at La Silla Observatory, combined with iÕ-band data obtained with MegaCam/MegaPrime at the CFHT. For the WFI BVR filters, we were able to use existing co-added images (B: 9,200 s, V: 8,700 s, R: 21,000 s) from a weak lensing follow-up of clusters detected in the Sunyaev-ZelÕdovich (SZ) effect. Observations spanning three campaigns between September 2000 and October 2011 were bias-subtracted and flat-fielded using the THELI processing pipeline\cite{erben05,schirmer13}. THELI also includes astrometric, relative and absolute photometric calibration. Finally, the CFHT iÕ-band data obtained in July 2009 were reduced using the CFHT-specific THELI adaptation developed and applied for the CFHTLenS project\cite{erben13}.
For all filters, the co-added images were post-processed, and saturated stars and otherwise unreliable image areas were masked out\cite{dietrich07}. Source catalogues were distilled from the co-added images using the weak lensing pipeline from ref 22. Because of the different field-of-view of the cameras involved ($34'\times34'$ for WFI versus $60'\times60'$ for CFHT MegaCam), it proved useful to adopt the following strategy: we measured source photometry in all three WFI passbands in one go, making use of the ``double detection'' mode in SEXTRACTOR\cite{bertin96}, with the deep R-band data as the detection image. In order to obtain consistent magnitudes, photometric quantities were measured after having matched the seeing in the other filters to the poorest seeing among them. A separate detection run was performed for the CFHT i'-data. The output catalogues were merged, identifying as the same object sources detecting in WFI and CFHT within 0.5 arcsec of each other, yielding a common photometric catalogue containing 37 WFI galaxies per square arcmin. Objects were categorized as stars or galaxies based on their apparent size and magnitude. \\

{\large \textbf{\emph{Lensing Analysis of HST and CFHT data}}}\\

\underline{\emph{Lensing Constraints :  HST field of view}}\\
The strong lensing constraints used to model the inner core of Abell 2744 consist of a set of 51 multiply-imaged systems (159 images\citebody{jauzac14b}). The weak lensing catalogue for the HST field of view was built following the methods described in ref 43, and the details of the Abell 2744 weak-lensing catalogue will be given elsewhere (M. J. et al., manuscript in preparation). Here we give a brief summary of the different steps. 

The weak lensing analysis is based on shape measurements in the Advanced Camera for Surveys (ACS)/F814W band. Following a method developed for the analysis of data obtained for the COSMOS survey\cite{leauthaud07}, the SEXTRACTOR photometry package\cite{bertin96} was used for the detection of the sources. The resulting catalogue was then cleaned by removing spurious sources, duplicate detections, and any sources in the vicinity of stars or saturated pixels. Finally, to overcome the pattern-dependent correlations introduced by the drizzling process between neighbouring pixels, we simply scaled up the noise level in each pixel\cite{leauthaud07} by the same constant $FA \approx0.316$\cite{casertano00}.

Since only galaxies behind the cluster are gravitationally lensed, the presence of cluster members dilutes the observed shear and reduces the significance of all quantities derived from it. Therefore, the identification and removal of the contaminating unlensed galaxies is crucial. Thanks to the HST data in three bands (F814W, F606W, and F435W), we identified the foreground galaxies and cluster members using a colour-colour diagram\citebody{jauzac14b}. The measure of galaxy shapes is done using the Rhodes-Refregier-Groth (RRG) method\cite{rhodes00}, adapted to multi-epoch images like the one coming from the HSTFF data of Abell 2744\cite{harvey15}. Finally, galaxies with ill-determined shape parameters were excluded, since these galaxies do not contribute significantly to the shear signal\cite{jauzac12}\citebody{jauzac14b}.\\

\underline{\emph{Lensing Constraints :  CFHT field of view}}\\
We employed the popular Kaiser-Squires-Broadhurst (KSB) method for galaxy shear measurement\cite{kaiser95}. We modelled the observed galaxy shape as a convolution of the (sheared) galaxy with the point spread function (PSF), which is itself modelled as a circular profile convolved with a small anisotropy. 
For the PSF modelling, we identified stars in the size-magnitude and $\mu_{\max}$-magnitude planes chip by chip\cite{shan12}, where $\mu_{\max}$ is the peak surface brightness. We then measured the Gaussian-weighted shape moments of the stars, and constructed their ellipticity. In addition to cuts in $\mu_{\max}$ and magnitude, we also excluded noisy outliers with signal-to-noise ratio SNR$<100$ or absolute ellipticity more than 2$\sigma$ away from the mean local value, and we iteratively removed objects very different from neighbouring stars. Having obtained our clean sample of stars, a second-order polynomial model in x and y was used to model the PSF across the field of view. The ellipticity of the PSF changes from its core to its wings. We measured the PSF shape using weight functions of different sizes and, when correcting each galaxy, used the weight function of the same size to measure the shapes of both the PSF and the galaxies. 
Background galaxies were selected with the magnitude cuts $20<i'<26$, size cuts $1.15r_{\rm PSF}<r_h<10$ pixel (where $r_h$ is the half-light radius and $r_{\rm PSF}$ is the size of the largest star), SNR$>$10 and SEXTRACTOR flag FLAGS=0. After masking and catalog cuts, the galaxy number density is $\sim10$ galaxies per square arcmin. We then measured the shapes of all the selected galaxies. Our implementation of KSB is based on the KSBf90 pipeline\cite{shan12}. The details of the calibration and systematic effects are shown and discussed elsewhere\cite{shan12}.
If the PSF anisotropy is small, the shear $\gamma$ can be recovered to first order from the observed ellipticity $e^{\rm obs}$ of each galaxy via
\begin{equation}\gamma=P_\gamma^{-1} \left(e^{\rm obs}-\frac{P^{\rm sm}}{P^{\rm sm*}}\right) e^* ,\end{equation}
\noindent where asterisks indicate quantities that should be measured from the PSF model interpolated to the position of the galaxy, $P^{\rm sm}$ is the smear polarizability, and $P_\gamma$ is the correction to the shear polarizability that includes smearing with the isotropic component of the PSF. The ellipticities were constructed from a combination of each objectÕs weighted quadrupole moments, and the other quantities involve higher order shape moments. All definitions are taken from \cite{luppino97}. Note that we approximate the matrix $P_\gamma$ by a scalar equal to half its trace. Since measurements of Tr $P_\gamma$  from individual galaxies are noisy, we fit it as a function of galaxy size and magnitude, which are more robustly observable galaxy properties\cite{shan12}.

The weight of the shear contribution from each galaxy is defined as
\begin{equation}w=\frac{P_\gamma^2}{\sigma_0^2 P_\gamma^2+\sigma_{e,i}^2 } ,\end{equation}
\noindent where $\sigma_{e,i}$ is the error for an individual galaxy obtained via the formula in Appendix A of \cite{hoekstra00}, and $\sigma_0\sim0.3$ is the dispersion of the intrinsic ellipticities of galaxies. With the help of the shear catalogue, we then estimated the total mass within the filaments. As the weak lensing effect is not very sensitive to the mass profile, we assumed a dual pseudo isothermal elliptical (dPIE) profile centred on the X-ray position to measure the total mass of the filament candidates using the parametric model-fitting algorithm LENSTOOL\cite{jullo07}.  As the weak lensing effect is not very sensitive to the mass profile, we also tested the accuracy of the derived masses by fitting again the shear profile with an elliptical Navarro-Frenk-White (NFW) profile with a concentration $c=1$. The measured masses are consistent within the uncertainties.\\

\underline{\emph{Lensing Mass Model}}\\
The mass model built for this analysis used strong and weak lensing constraints, combining parametric and non-parametric approaches to model the global mass distribution\cite{jauzac15}. The details of the mass modeling will be given elsewhere (M.J. et al., manuscript in preparation). We kept the parameters of the model built for the strong lensing analysis of Abell 2744 fixed to their best-fit values, and we modelled the surrounding mass distribution using a multi-scale grid drawn from a prior light distribution of the cluster using the WFI multi-band photometric catalogue. The nodes of the grid model were parameterized using Radial Basis Functions (RBFs\cite{jullo14}). This allowed us to appropriately weight the strong lensing constraints without taking them twice into account\cite{jauzac15}. 

The strong lensing parametric model was composed of two cluster-scale halos. The multi-scale grid was composed of 10,282 RBFs, for which only the amplitude was left free while fitting. To the 733 cluster members identified in the HST fields-of-view, we added 1457 cluster members identified using a standard colour-magnitude selection using B, V and R bands coming from WFI observations to identify the red-sequence galaxies of the cluster. Galaxy-scale haloes were modelled as RBFs, using dPIE potentials. The resulting mass map is shown by the white contours in Figure 3.

We sampled the parameter space in LENSTOOL using the Bayessys Library implemented in LENSTOOL\cite{jullo07}. The objective function is a standard likelihood function in which noise is assumed to be Gaussian. LENSTOOL returns a large number of MCMC samples, from which we estimate mean values and uncertainties in the mass density field. In Extended Data Fig. 5 we show the radial surface mass density profile for the cluster average compared to the sectors encompassing the filaments (same as for Extended Data Fig. 1). An excess lensing signal is observed in the direction of the filaments compared to the radial average. The masses obtained using this technique are given in Table 1. In Extended Data Table 2 we show the masses and SNRs obtained using this method (hybrid LENSTOOL) and the direct inversion method described above (KSB) for the various filaments. The results of the two methods agree within the uncertainties. The differences observed between one method and the other allow us to quantify the level of systematic uncertainties associated with the lensing reconstruction using the existing data.

\begin{figure}
\resizebox{\hsize}{!}{\includegraphics{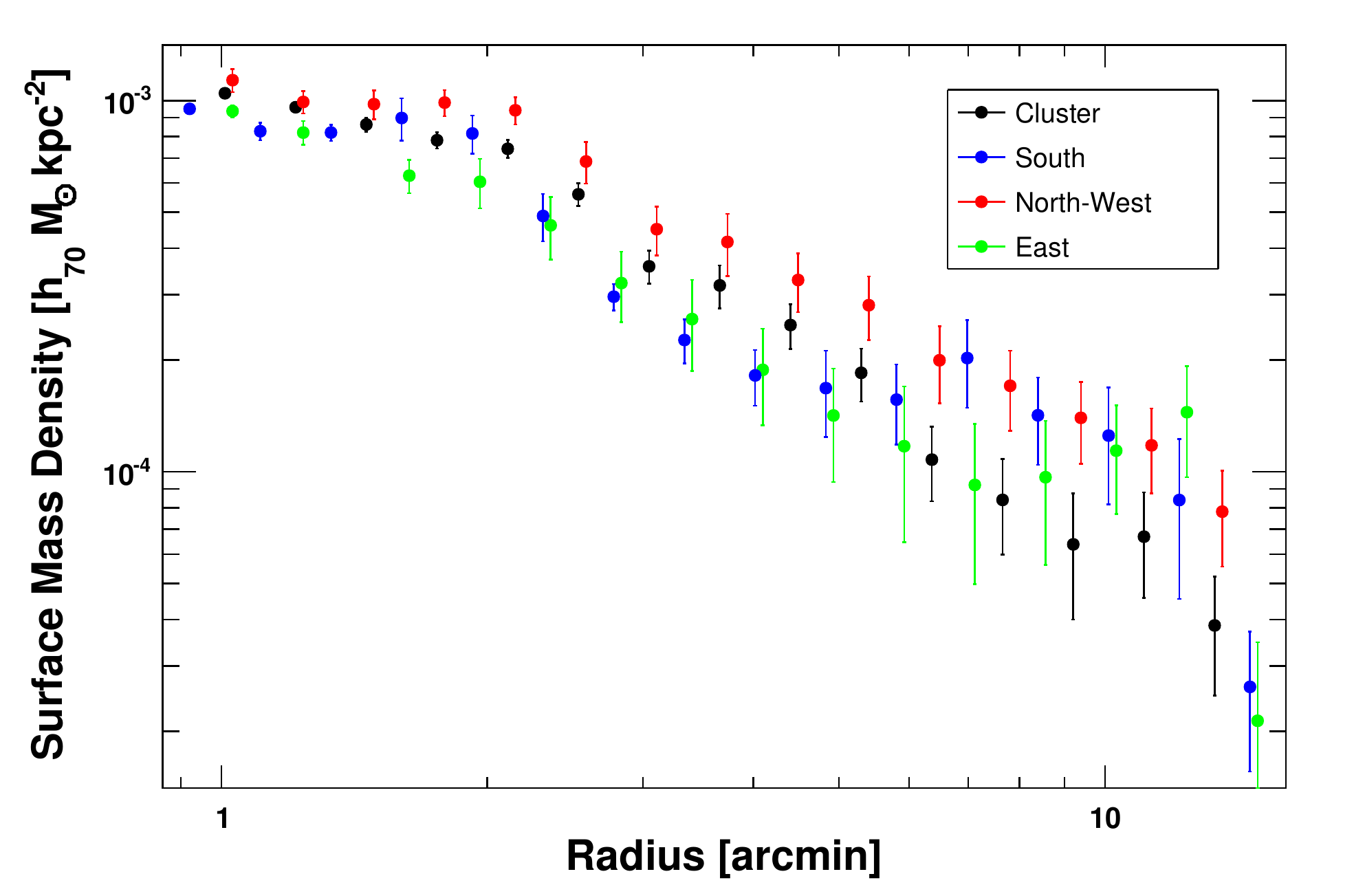}}
\textbf{Extended Data Figure 5: Radial mass profiles in the filaments and in the cluster.} Shown are surface mass density profiles obtained from combined strong and weak lensing. The black curve shows the cluster average, compared to the profiles obtained in the direction of the filaments (same as Extended Data Fig. 1).
\end{figure}

\begin{table*}
\begin{center}
\begin{tabular}{ccccc}
\hline
Region & $M_{\rm HLT}$ & S/N & $M_{\rm KSB}$ & S/N\\
 & [$h_{70}^{-1}M_\odot$] &  & [$h_{70}^{-1}M_\odot$] &  \\
\hline
\hline
E & $(7.9\pm2.8)\times10^{13}$ & 3.1 & $(4.4\pm3.1)\times10^{13}$ & 2.1\\
S & $(9.5\pm2.4)\times10^{13}$ & 6.8 & $(4.0\pm2.4)\times10^{13}$ & 2.3\\
SW & $(4.8\pm1.7)\times10^{13}$ & 3.1 & $(2.2\pm1.6)\times10^{13}$ & 2.8\\
NW1 & $(9.5\pm2.7)\times10^{13}$ & 5.2 & $(6.9\pm3.0)\times10^{13}$ & 2.2\\ 
NW2 & $(1.2\pm0.3)\times10^{14}$ & 3.3 & $(2.2\pm1.0)\times10^{14}$ & 2.6\\
 \hline
\end{tabular}
\end{center}
\textbf{Extended Data Table 2: Mass of the filaments.} Comparison of weak-lensing masses for the filaments for the two methods used here: the direct inversion method (KSB) and the grid-based multi-scale approach (hybrid LENSTOOL, HLT).
\end{table*}

{\small
\renewcommand{\refname}{References}
\bibliographystyle{naturemag}
\bibliography{nature}}

\newpage
\vspace{0.3cm}
\noindent {\large \textbf{Sample size.}}\\ 

No statistical methods were used to predetermine sample size.

\vspace{0.3cm}
\noindent {\large \textbf{Code Availability}}\\

$\bullet$ The PROFFIT code for X-ray surface brightness analysis is available at: \\
\url{http://www.isdc.unige.ch/~deckert/newsite/Proffit.html} \\
$\bullet$ The THELI data reduction scheme for CFHT and ESO/WFI data can be downloaded at:\\
\url{https://www.astro.uni-bonn.de/theli/}\\
$\bullet$ The gravitational lensing code LENSTOOL can be found at:\\
\url{http://projets.lam.fr/projects/lenstool/wiki}\\
$\bullet$ The KSBf90 code used for weak lensing is available at:\\
\url{http://www.roe.ac.uk/~heymans/KSBf90/Home.html}

\end{document}